\documentclass{article}
\usepackage{amsmath,graphicx,amssymb,amsfonts,latexsym,color}
\usepackage{bm}
\usepackage{multirow}
\usepackage{tikz}
\usepackage[round]{natbib} 
\usepackage[ruled,vlined]{algorithm2e}
\usepackage{subcaption}
\usepackage{textcomp}
\usepackage{hyperref}

\begin{document}
\title{A flexible model-based framework for robust estimation of mutational signatures}
\date{\today}
\author{Ragnhild Laursen$^{1}$, Lasse Maretty$^{2}$ and Asger Hobolth$^{3}$ \\
\small
1. Department of Mathematics, Aarhus University. 
Email: ragnhild@math.au.dk \\
\small
2. Department of Clinical Medicine and Bioinformatics Research Center, Aarhus University. \\
\small
Email:  lasse.maretty@clin.au.dk \\
\small
3. Department of Mathematics, Aarhus University. 
Email: asger@math.au.dk}
\maketitle
\section*{Abstract}
Somatic mutations in cancer can be viewed as a mixture distribution of several mutational signatures, which can be inferred using non-negative matrix factorization (NMF). Mutational signatures have previously been parametrized using either simple mono-nucleotide interaction models or general tri-nucleotide interaction models. We describe a flexible and novel framework for identifying biologically plausible parametrizations of mutational signatures, and in particular for estimating  di-nucleotide interaction models. The estimation procedure is based on the expectation--maximization (EM) algorithm and regression in the log-linear quasi--Poisson model. We show that di-nucleotide interaction signatures are statistically stable and sufficiently complex to fit the mutational patterns. Di-nucleotide interaction signatures often strike the right balance between appropriately fitting the data and avoiding over-fitting. They provide a better fit to data and are biologically more plausible than mono-nucleotide interaction signatures, and the parametrization is more stable than the parameter-rich tri-nucleotide interaction signatures. We illustrate our framework on three data sets of somatic mutation counts from cancer patients. 
\vspace{2mm} \newline  
\noindent {\em Key words:}
Cancer genomics, expectation-maximization (EM) algorithm, interaction terms, mutation counts, mutational signatures,  Non-negative Matrix Factorization (NMF),  Poisson regression.
\vspace{2mm} \newline
\noindent {\em AMS classification:}
Primary: 62 (Statistics), Secondary: 62F10 (Point estimation), 62F30 (Parametric inference under constraints), 62H12 (Estimation in multivariate analysis), 62P10 (Applications of statistics to biology and medical sciences), 68T05 (Learning and adaptive systems in artificial intelligence), 92B20 (Neural networks in biological studies).  
\section{Introduction}
The mutation rate at a particular site in the genome often depends on both the left and right flanking nucleotides. \cite{Hwang13994} analysed a 1.7 mega-base alignment of 19 mammalian species, and perhaps the most striking observation was a much elevated mutation rate for $C>T$ mutations when the right flanking nucleotide is a $G$. The elevated rate reflects deamination of methyl cytosine. The $CG$-methylation-deamination process was the main focus in the neighbour-dependent models described in \cite{ArndtBurgeHwa2003} and \cite{Hobolth2008}. Furthermore, longer contextual patterns are also known to impact mutation rates \citep{lindberg2019intragenomic}.  

Analyses of somatic mutations in cancer patients have increased our basic understanding of the mutational processes operating in human cancer (\cite{alexandrov2020}). For example, mutational signatures from tobacco smoking (\cite{Alexandrov2016}) and UV-light (e.g. \cite{Shen2020}) have been identified. Furthermore, mutational signatures can be used as biomarkers for deciding the diagnosis and treatment of cancer patients (\cite{Nik-Zainal2017}).   
A simple parametrization of mutational signatures is essential to achieve statistically stable estimation, easier interpretation of signatures, and the possibility of including more flanking nucleotides than just the nearest neighbors. 

The purpose of this paper is to formulate a flexible framework for parmetrizing mutational signatures. The mutational signatures from \cite{alexandrov2013} and \cite{shiraishi2015} constitute two extremes in our framework. We view signatures as a composition of interactions between the mutation type and the flanking nucleotides. In this context, the general model from \cite{alexandrov2013} with 96 mutation types include all tri-nucleotide interaction terms, and the independence model from \cite{shiraishi2015} has no interaction terms between the mutation and the flanking nucleotides i.e. mono-nucleotide interaction terms. We propose a model that reaches the middle-ground between the complex model of \cite{alexandrov2013} and the simple model of \cite{shiraishi2015}. Our model includes di--nucleotide interaction terms between the mutation type and flanking nucleotides. We also investigate combinations of the various parametrizations of mutational signatures. 

We analyse three data sets of somatic mutations in cancer patients. The first two data sets are from breast cancer patients with 96 mutation types. We analyse the 21 breast cancer patients from \cite{alexandrov2013} and the 214 breast cancer patients from \cite{alexandrov2020}. We refer to these data sets as BRCA21 and BRCA214. 
We show that mono-nucleotide signatures give a poor fit to the data and tri-nucleotide signatures over-fit the data, while di--nucleotide signatures strike the right balance between the two. We also estimate a flexible mixture model where each signature can be either mono-nucleotide, di-nucleotide or tri-nucleotide. In the mixture model we find evidence of mutational signatures that exhibit either tri--nucleotide, di--nucleotide and mono--nucleotide interaction terms.

The third data set is from urothelial carcinoma of the upper urinary tract \citep{Hoang2013} from 26 patients with 1536 mutation types. These mutation types include two flanking nucleotides to each sides of the base mutation. This data was also used by \cite{shiraishi2015}, and we refer to the data as UCUT26. We find that the di-nucleotide interaction models fit the data substantially better than the mono-nucleotide models and are statistically much more stable than the full penta-nucleotide model.

Our analyses thus validate the relevance of our flexible framework for mutational signatures. The di–nucleotide signatures provide a better fit to the data and are biologically more plausible than mono-nucleotide signatures, and the parametrization is more stable than the parameter-rich higher-order signatures that include all interaction terms. 
 
Our paper is organized as follows. In Section~\ref{sec:Param} we describe non-negative matrix factorization and parametrization of a mutational signature in terms of interactions between the nucleotides in the mutation type. In Section~\ref{sec:Analysis} we analyse the BRCA21, BRCA214 and UCUT26 data sets. Maximum likelihood estimation is carried out using a novel combination of the expectation-maximization algorithm (\cite{DempsterLairdRubin1977}) and regression in the quasi-Poisson model (e.g. \cite{McCullaghNelder1989}), and is described in detail in Section~\ref{sec:Methods}. The paper ends with a general discussion about parametrization and model selection for mutational signatures. The data and code for reproducing the results and figures are available at \url{https://github.com/ragnhildlaursen/paramNMF_ms}. 
\section{Determining the mutational signatures} \label{sec:Param}
Mutational signatures are derived from mutational counts using an unsupervised method called non-negative matrix factorization (NMF). In this section we first explain NMF in general terms and afterwards how parameterization of the mutational signatures is included in the method.
\subsection{Non-negative matrix factorization} \label{sec:nmf}
Given a data matrix $V \in \mathbb{N}_+^{N \times T}$, the main aim of non-negative matrix factorization (NMF) is to find a factorization $WH$, where the product of the non-negative exposure (sometimes also called weight or loading) matrix $W \in \mathbb{R}_+^{N \times K}$ and the non-negative signature matrix $H \in \mathbb{R}_+^{K \times T}$ provide a good approximation of the data matrix, i.e.
\begin{equation}
    V \approx WH. 
\end{equation}
In our application $N$ is the number of cancer patients, $T$ is the number of mutation types, and each entry $V_{nt}$ is the total number of somatic cancer mutations of type $t$ in patient $n$. The weight matrix $W$ is of size $N\times K$ and $H$ is a matrix of mutational signatures of size $K \times T$, where they both only include non-negative entries. Each of the $K$ signatures is a discrete probability distribution of length $T$, i.e. has $T-1$ free non-negative parameters that sum to at most one. The rank $K$ of the factorization is most often one or more magnitudes smaller than the minimum of $N$ and $T$.

For mutational count data it is natural to assume that each entry is Poisson distributed
\begin{equation}
    V_{nt} \sim \text{Pois} \big( (WH)_{nt} \big), \quad n = 1, \dots N, \ t = 1, \dots , T.
    \label{eq:poisson}
\end{equation}
The data log-likelihood is, up to an additive constant, given by
\begin{equation}
    \ell(W,H;V) = 
    \sum_{n=1}^N \sum_{t=1}^T 
    \Big\{ V_{nt} \log \big( (WH)_{nt} \big) - (WH)_{nt} \Big\}, 
    \label{eq:datalik}
\end{equation}
and we determine $W$ and $H$ by maximizing the data log-likelihood. The details are provided in Section~\ref{sec:Methods}. Maximization of the data log-likelihood is identical to minimizing the generalized Kullback-Leibler (GKL) divergence
\begin{equation}
    {\rm GKL}=
    \sum_{n=1}^N \sum_{t=1}^T 
    \Big\{ V_{nt} \log V_{nt}-V_{nt}\log \big( (WH)_{nt} \big)-V_{nt} + (WH)_{nt} \Big\}.
    \label{GKL}
\end{equation}
This follows as the negative data log-likelihood is proportional to the GKL up to an additive constant. 
The factorization is clearly not unique up to permutation and scaling. Indeed, if $W$ and $H$ are non-negative and $A$ is a $K\times K$ permutation matrix, we have that $WA$ and $A^{-1}H$ are non-negative and $WH=W(AA^{-1})H=(WA)(A^{-1}H)$. The permutation issue is taken into account by a potential re-ordering of the mutational signatures and their corresponding weights. If $A$ is a diagonal matrix with positive entries we also have that $WA$ and $A^{-1}H$ are non-negative and $WH=(WA)(A^{-1}H)$. The scaling issue can be solved by normalizing the signatures in $H$ such that they sum to one, i.e. by choosing $A={\rm diag}(d_1,\ldots,d_K)$ as the diagonal matrix with entries $d_k=\sum_{t=1}^T H_{kt}$, $k=1,\ldots,K$, on the diagonal. We refer to \cite{LaursenHobolth2021} for a general discussion of the NMF non-uniqueness problem and a general procedure to determine the set of feasible solutions.  

In general, the number of observations is $N\cdot T$ and the number of parameters is $N\cdot K$ for the weight matrix and $K\cdot (T-1)$ for the signature matrix. 
With $N=21$ patients and $K=4$ signatures the number of observations $N\cdot T=21\cdot 96=2016$ are estimated using $N\cdot K +K\cdot (T-1)=21\cdot 4+4\cdot 95=84+380=464$ free parameters. Thus, in general, this approach has a large number of free parameters compared to the size of the data matrix. These considerations suggest that parametrizing a mutational signature is fruitful.
\subsection{Parametrization of a mutational signature}
We parametrize each mutational signature $h= (h_1, \dots, h_T)$ by the mutation type as a function of the base mutation $M$, the flanking left base $L$ and the flanking right base $R$. The number of mutations is 12 without strand-symmetry, and 6 with strand-symmetry. Each flanking nucleotides can be one of the four types $A, \ C, \ G$ or $T$. The different factors are thus the left neighbour $L$ (4 categories), the right neighbour $R$ (4 categories) and the mutation type $M$ (6 or 12 categories). In all of the following we assume strand-symmetry, so that $M$ has 6 categories.

We model the mutational signatures with a log-linear parametrization given by
\begin{equation}
  h_t = \frac{\exp( (X \beta)_t )}{\sum_{t=1}^T \exp( (X \beta)_t )},
  \quad t=1,\ldots,T,
  \label{logH}
\end{equation}
where $X$ has dimension $T \times S$ and is the design matrix that describe the common factors among the different mutation types and $\beta \in \mathbb{R}^S$ is a vector of $S$ parameters for the different factors. 
\subsubsection{One flanking nucleotide at each side of the mutation}
The mutational signature $h$ with one flanking nucleotide at each side is a vector of length $T=4\cdot 6 \cdot 4=96$ indexed by ${\ell m r}$. 
Following classical factorial analysis of variance we specify the general tri-nucleotide interaction model from \cite{alexandrov2013} by $L\times M \times R$. The model can be written as
\begin{eqnarray}
  h_{\ell mr} = 
  \frac{\exp(\beta^{L\times M\times R}_{\ell m r})}
       {\sum_{\ell \in L} \sum_{m \in M} \sum_{r \in R} 
       \exp(\beta^{L\times M\times R}_{\ell m r})},
\label{TriNucleotide}
\end{eqnarray}
where $m$ describes the six base mutation, and $\ell$ and $r$ describe the four possible flanking nucleotides to the left or right of the base mutation. This gives $S=T=4 \cdot 6 \cdot 4 = 96$ different parameters in the $\beta$ vector and $X = I_T$ is the $T\times T$ identity matrix in the general formulation \eqref{logH}.

The mono-nucleotide interaction model $L+M+R$ of \cite{shiraishi2015} takes the form 
\begin{eqnarray}
  h_{\ell mr} = 
  \frac{\exp(\beta^M_{m} + \beta^L_{\ell} + \beta^R_{r})}
       {\sum_{\ell \in L} \sum_{m \in M} \sum_{r \in R} 
        \exp(\beta^M_{m} + \beta^L_{\ell} + \beta^R_{r})}.
\label{MonoNucleotide}
\end{eqnarray}
In order to avoid confounding we define $\beta_A^R=\beta_A^L=0$. 
Therefore, we have $S=3 + 6 + 3 = 12$ remaining parameters in the $\beta$ vector, which is a substantial reduction from the original model with $96$ parameters. 
The corresponding $96\times 12$ design matrix $X$ takes the form
\begin{equation}
X =
\begin{tabular}{ccccccc|ccc|cccccccc}
        & \multicolumn{6}{c|}{ \textbf{M}utation} 
        & \multicolumn{3}{c|}{ \textbf{L}eft base}  
        & \multicolumn{3}{c}{  \textbf{R}ight base} 
        & \\
        & \scriptsize $C > A$ & \scriptsize $C > G$ & \scriptsize $C > T$ 
        & \scriptsize $T > A$ & \scriptsize $T > C$ & \scriptsize $T > G$ 
        & \scriptsize $C$    & \scriptsize $G$    & \scriptsize $T$  
        & \scriptsize $C$    & \scriptsize $G$    & \scriptsize $T$    
        & \\
        \scriptsize $A [ C > A ] A$ 
        & \hspace{-0.9cm} 
          \multirow{5}{*}{$\left. \rule{0cm}{1.5cm} \right($ } 
          \hspace{0.1cm} 
        1 & 0 & 0 & 0 & 0 & 0 & 0 & 0 & 0 & 0 & 0 & 0 & 
          \hspace{-0.5cm} 
          \multirow{5}{*}{$\left. \rule{0cm}{1.5cm} \right)$ } \\
        \scriptsize $A [ C > A ] C$ 
        & 1 & 0 & 0 & 0 & 0 & 0 & 0 & 0 & 0 & 1 & 0 & 0  \\
        \scriptsize $A [ C > A ] G$ 
        & 1 & 0 & 0 & 0 & 0 & 0 & 0 & 0 & 0 & 0 & 1 & 0  \\
        \vdots & \multicolumn{6}{c|}{\vdots} & \multicolumn{3}{c|}{\vdots}  & \multicolumn{3}{c}{\vdots} & \\
        \scriptsize $T [ T > G ] T$ 
        & 0 & 0 & 0 & 0 & 0 & 1 & 0 & 0 & 1 & 0 & 0 & 1    
\end{tabular}.
\vspace{1em}
\end{equation}

We propose the di-nucleotide interaction signature $L\times M+M\times R$ given by
\begin{eqnarray}
  h_{\ell mr} = 
  \frac{\exp(\beta_m^{M}+\beta_{\ell m}^{L\times M} + \beta_{m r}^{M \times R})}
       {\sum_{\ell \in L} \sum_{m \in M} \sum_{r \in R}
        \exp(\beta_m^{M}+\beta_{\ell m}^{L\times M} + \beta_{m r}^{M \times R})}.
\label{DiNucletide}
\end{eqnarray}
In order to avoid confounding we define $\beta_{Am}^{L\times M}=\beta_{mA}^{M \times R}=0$ for all the six possible base mutations $m \in \{ { \scriptsize C > A, C > G, C>T, T>A, T>C, T > G} \}$. This signature therefore has a total of $S=3 \cdot 6 + 6 + 3 \cdot 6 = 42$ parameters and is a biologically plausible alternative between the simple mono-nucleotide multiplicative signature of \cite{shiraishi2015} and the complex tri-nucleotide interaction signature of \cite{alexandrov2013}. From the mutational pattern of spontaneous cytosine deamination in CpG contexts, we know that some processes are dependent on only one neighbouring nucleotide \cite{ArndtBurgeHwa2003}. A summary of the three models is provided in Table~\ref{tab:ModelOne}. Results for these models are shown for the breast cancer patients in Section~\ref{sec:brca21} and Section~\ref{sec:brca214}.

\begin{table}[h!]
\centering
\begin{tabular}{lcrl}
   \multicolumn{4}{c}{One flanking nucleotide at each side} \\ \hline
  Signature & Factorization & Number of parameters & Key reference  \\
  \hline \hline
  Mono-nucleotide & $L+M+R$  & $3+6+3=6+3\cdot2=12$  &    \cite{shiraishi2015} \\
  Di-nucleotide & $L\times M+M\times R$  & $3\cdot6+6+3\cdot 6=6+18\cdot2=42$ & Our proposed model \\
  Tri--nucleotide & $L\times M \times R$ & $4\cdot6\cdot4=6\cdot4^2=96$ & \cite{alexandrov2013}    
\end{tabular} 
\caption{Parametrizations of a mutational signature with one flanking nucleotide at each side and increasing complexity.}
\label{tab:ModelOne}
\end{table}
\subsubsection{Two flanking nucleotides at each side of the mutation}
\cite{shiraishi2015} considers higher-order context dependencies where the mutation types include four flanking bases, which gives five different factors $L_2, L_1, M, R_1$ and $R_2$. The number of mutation types in this case is $T=4^2\cdot 6 \cdot 4^2=6\cdot 4^4=1536$ and the number of parameters in the mono-nucleotide model with two flanking neighbours on each side of the mutation is $3+3+6+3+3=6+3\cdot(2\cdot2)=18$.

In general $T=6\cdot 4^{2n}$ when $n$ bases are considered upstream and downstream of the mutated site, and the number of mutation types $T$ (and signature parameters in the general model) thus increases exponentially with the number of neighbouring nucleotides.     
There are $6+3(2n)=6(1+n)$ parameters in the mono-nucleotide model, i.e. a linear increase in the number of parameters. In this paper we introduce di-nucleotide models that include interactions between neighbors given by $L_1\times M+M\times R_1+\sum_{i=1}^{n-1} (L_{i+1}\times L_i+R_i\times R_{i+1})$. This model results in $42+12 \cdot 2 \cdot (n-1) = 18 + 24 \cdot n$ parameters. Thus, our di-nucleotide signatures are also linear in the number of flanking nucleotides. 

\begin{figure}[h!]
    \centering
    \includegraphics[width = \textwidth]{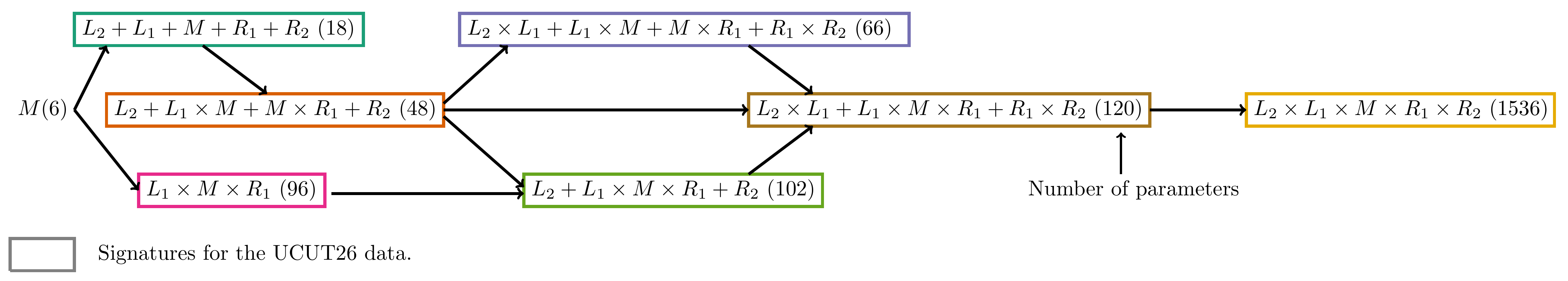} \\
    \caption{Factor diagram for the signatures used for the UCUT26 data set. The diagram shows the number of parameters for each signature and how the signatures are nested in each other.}
    \label{fig:factordiagram}
\end{figure}

Our framework is very flexible, and we are able to analyse combinations of mono-, di- and tri-nucleotide interaction terms within a signature. We consider the signatures $L_2+L_1\times M+M \times R_1+R_2$, $L_2+L_1 \times M \times R_1+R_2$ and $L_2 \times L_1 +L_1\times M \times R_1+ R_1 \times R_2$. These three are combinations of mono-, di- and tri-nucleotide intereactions. See~Table~\ref{tab:ModelTwo} and Figure~\ref{fig:factordiagram} for an overview of the signatures with two flanking nucleotides at each side and how they are nested in each other. Further results for these models on the UCUT data are seen in Section~\ref{sec:ucut}.   
\begin{table}[h!]
\centering
\begin{tabular}{lcr}
  \multicolumn{3}{c}{Two flanking nucleotides at each side}  \\ \hline
 Signature & Factorization & Number of parameters \\ \hline \hline
 Mono-nucleotide & $L_2+L_1+M+R_1+R_2$ & $6+3\cdot 4=18$ \\
 Di-nucleotide & $L_2 \times L_1+L_1\times M+M\times R_1+R_1\times R_2$ &
 $42+12\cdot 2=66$ \\
 Tri-nucleotide & $L_1 \times M \times R_1$ &
 $6 \cdot 4^2 =96$ \\
 Penta-nucleotide & $L_2 \times L_1 \times M \times R_1 \times R_2$ & $6\cdot 4^4=1536$ \\ \hline
 Di- and mono-nucleotide & $L_2+L_1\times M+M \times R_1+R_2$ & $42+3\cdot 2=48$ \\ 
 Tri- and mono-nucleotide & $L_2+L_1\times M \times R_1+R_2$ & $96+3\cdot 2=102$ \\ 
 Di- and tri-nucleotide & $L_2 \times L_1 +L_1\times M \times R_1+ R_1 \times R_2$ & $96+12\cdot 2=120$
\end{tabular}
\caption{Parametrizations of a mutational signature with two flanking nucleotides at each side. We consider two categories of di-nucleotide interaction models. The first category has interaction between the flanking nucleotide and the mutation. The second category has interaction between the two nearest neighbours.}
\label{tab:ModelTwo}
\end{table}

\section{Data Analyses} \label{sec:Analysis} 
In this section we analyse the BRCA21, BRCA214 and UCUT26 data sets. We compare the various parametrizations and use parametric bootstrap and downsampling to investigate statistical robustness and stability of the signatures.

For the BRCA21 and BRCA214 data sets we follow \cite{alexandrov2013} and fix the number of signatures at $K=4$. Apart from the three models where all four signatures are parametrized the same way, we also consider a mixture parametrization of the signature matrix where each signature can be any of the three parametrizations in Table~\ref{tab:ModelOne}. We could investigate $3^4=91$ models, but recall from Section~\ref{sec:nmf} that the models are only identifiable up to permutation; this results in 15 different models.

For the UCUT26 with two flanking nucleotides at each side of the mutation we follow \cite{shiraishi2015} and fix the number of signatures to $K=2$. We could consider all combinations of the seven parametrizations in Table~\ref{tab:ModelTwo}, which results in ${6 \choose 2}+6=28$ different models. We have chosen to include 16 of the most relevant combinations for clarity. 

The best model can be determined by several methods that are balancing between a good fit to the data and avoiding over-fitting, and the choice depends on the application of the model (e.g. \cite{Shmueli2010}). In this paper we use the Bayesian Information Criteria (BIC) given by
\begin{equation*}
    {\rm BIC} = n_{\rm prm} \log n_{\rm obs}  - 2 \ell(W,H;V) 
              \equiv n_{\rm prm} \log n_{\rm obs}  + 2 {\rm GKL},
\end{equation*}
where $n_{\rm prm}$ is the number of parameters, $n_{\rm obs}$ is the number of observations, $\ell(W,H;V)$ is the log-likelihood function from (\ref{eq:datalik}), ${\rm GKL}$ is the generalized Kullback-Leibler divergence from (\ref{GKL}), and $\equiv$ means that the statement is true up to an additive constant.
Appropriate models have a small BIC because they represent a good balance between model complexity (measured in terms of the number of parameters) and goodness of fit (measured in terms of the negative log-likelihood).
\subsection{Analysis of BRCA21 data} \label{sec:brca21}
Recall that the BRCA21 breast cancer data set has $T=96$ mutation types and $N=21$. 
The number of observations for the data set is $n_{\rm obs}=T\cdot N=96\cdot 21=2016$.
We firstly consider the three models where all four signatures are either mono-nucleotide, di-nucleotide or tri-nucleotide. The number of parameters $n_{\rm prm}$, penalization term for number of parameters $n_{\rm prm}\log n_{\rm obs}$, GKL and difference between the BIC values and the smallest for each of the three models are summarized in Table~\ref{BRCA21tbl}, together with the mixture model with the smallest BIC. The GKL difference between the mono-nucleotide and di-nucleotide model is very large which means that the fit to the data is poor for the mono-nucleotide model.
\begin{table}[h!]
\centering
\resizebox{\textwidth}{!}{%
\begin{tabular}{lrrrr}
\hline
  & Number of parameters & Model complexity & Fit to data & Model selection \\ 
 Model for signatures & $n_{\rm prm}$ & $n_{\rm prm}\log n_{\rm obs}$ & ${\rm GKL}$  & $ \triangle {\rm BIC}$ \\ \hline \hline
 Sole mono-nucleotide & $4 \cdot 12=48$ & 365 & 4647 & 4487 \\
 Sole di-nucleotide & $4 \cdot 42=168$ & 1278 & 2059 & 224  \\
 Sole tri-nucleotide & $4 \cdot 96=384$ & 2922 & 1472 & 694 \\ \hline
 Mixture & $12+42+42+96=192$ & 1461 & 1856 & 0 \\
\end{tabular}
}
\caption{Summary statistics for the three basic models for the BRCA21 data where all signatures are the same, and a flexible mixture model where signatures can vary. The mixture model consists of one mono-, two di-, and one tri-nucleotide interaction signature, and has the smallest BIC. We have $K=4$ and the number of observations is $n_{\rm obs}=T\cdot N=2016$.}
\label{BRCA21tbl}
\end{table}

\begin{figure}[h!]
    \centering
    \includegraphics[width = 0.8\textwidth]{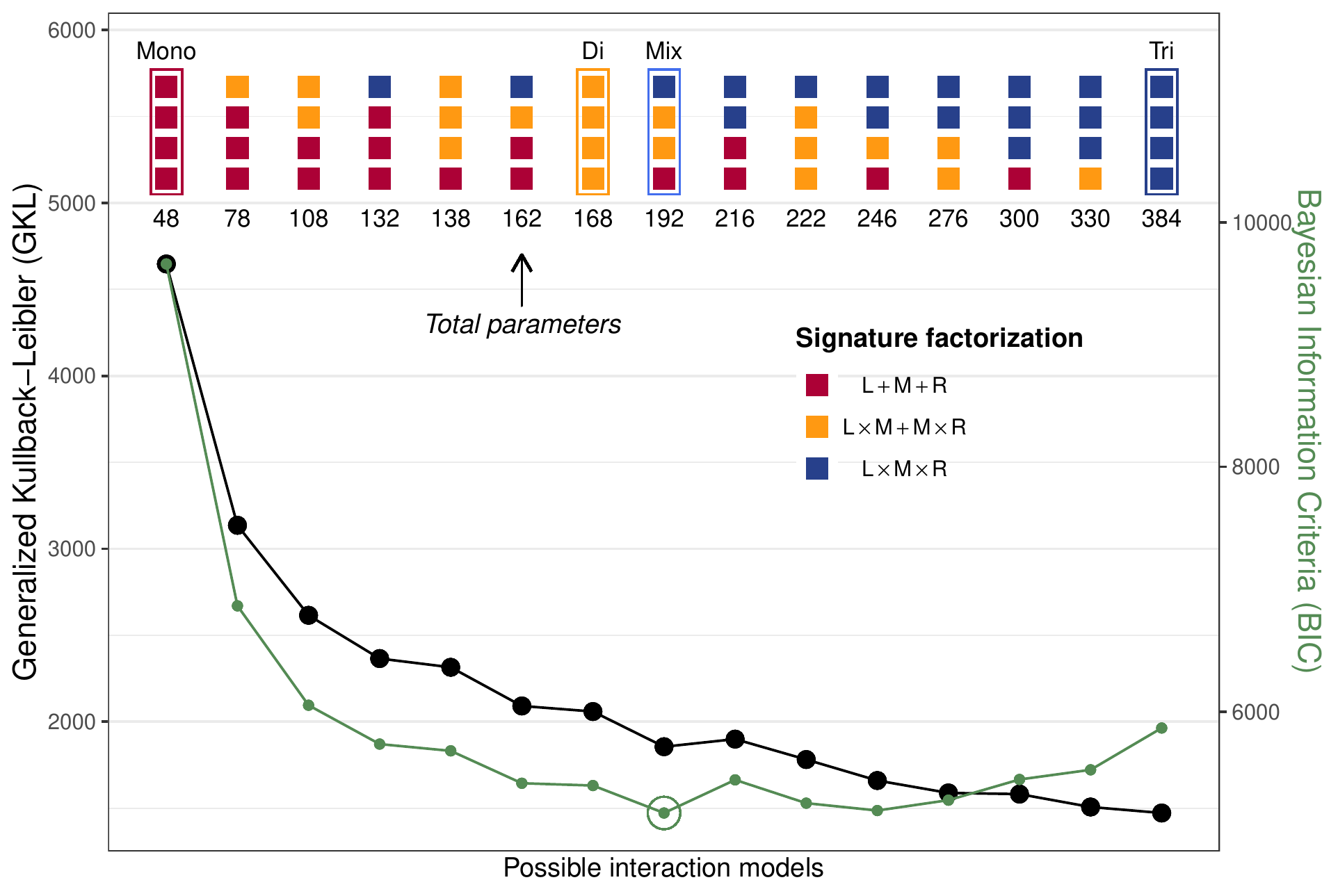} 
    \caption{Fit of 21 breast cancer patients for all possible interaction models. The Generalized Kullback-Leibler (GKL) and Bayesian Information Criteria (BIC) for all 15 models with 4 signatures. The models are ordered according to the total number of parameters for the 4 signatures; e.g. $4\cdot 12=48$ for the sole mono-nucleotide model and $4\cdot 96=384$ for the sole tri-nucleotide model. The model with the smallest BIC is indicated, and consists of one tri-nucleotide signature, two di-nucleotide signatures and one mono-nucleotide signature. }
    \label{fig:BRCA21models}
\end{figure}

We allow a flexible parametrization of type $L\times M \times R$, $L\times M+M\times R$, and $L+M+R$ for each of the $K=4$ signatures. For the 15 models, Figure \ref{fig:BRCA21models} shows the Generalized Kullback-Leibler divergence (GKL) and the Bayesian Information Criteria (BIC). The models are ordered according to the number of free parameters. The EM-algorithm can get stuck in local minima, so we start the algorithm by running 500 different initializations for 500 iterations and identify the minimum. From that minimum we then continue iterating until convergence. This procedure of starting the algorithm multiple times and running for a few iterations is recommended by \cite{biernacki2003choosing}. \cite{biernacki2003choosing} testing many different ways of running the EM-algorithm to escape local maxima and identify the global maximum likelihood value. 

We observe a steep decrease in GKL when the mono--nucleotide assumption is relaxed, and one or more signatures are allowed to contain di--nucleotide or even tri--nucleotide interactions. This shows that only applying mono--nucleotide signatures is biologically too restrictive.  
The mixture model with the smallest BIC (Mix in Figure~\ref{fig:BRCA21models}) has one mono--nucleotide signature, two di--nucleotide signatures and one tri-nucleotide interaction signature. Neither the fully independent model (Mono in Figure~\ref{fig:BRCA21models}) nor the fully general (Tri in Figure~\ref{fig:BRCA21models}) have a small BIC compared to the other models. The fit to the data is too poor for the independent model, and the general model has too many free parameters. The model with four di-nucleotide signatures (Di in Figure~\ref{fig:BRCA21models}) has a much smaller BIC than both the fully independent and fully general model. We conclude that a flexible parametrization of each signature allows us to obtain a good balance between a too constrained model with a poor fit to the data and a very parameter--rich model with a good fit to data.   

In order to investigate the statistical stability of the signatures we use parametric bootstrap. For a given model with an estimate of the count matrix $\hat{W} \hat{H}$ we simulate 50 data sets from the Poisson model~\eqref{eq:poisson}. For each of the simulated data sets we re-estimate the exposures and signatures and use cosine similarity to investigate how close the re-estimated signatures are to the true signatures under the specific model.  

\begin{figure}[b!]
 \centering
 \begin{tabular}{l}
      \noindent {\bf A. Inferred signatures for the BRCA21 data set} \vspace{1mm} \\
    \includegraphics[width = 0.9\textwidth]{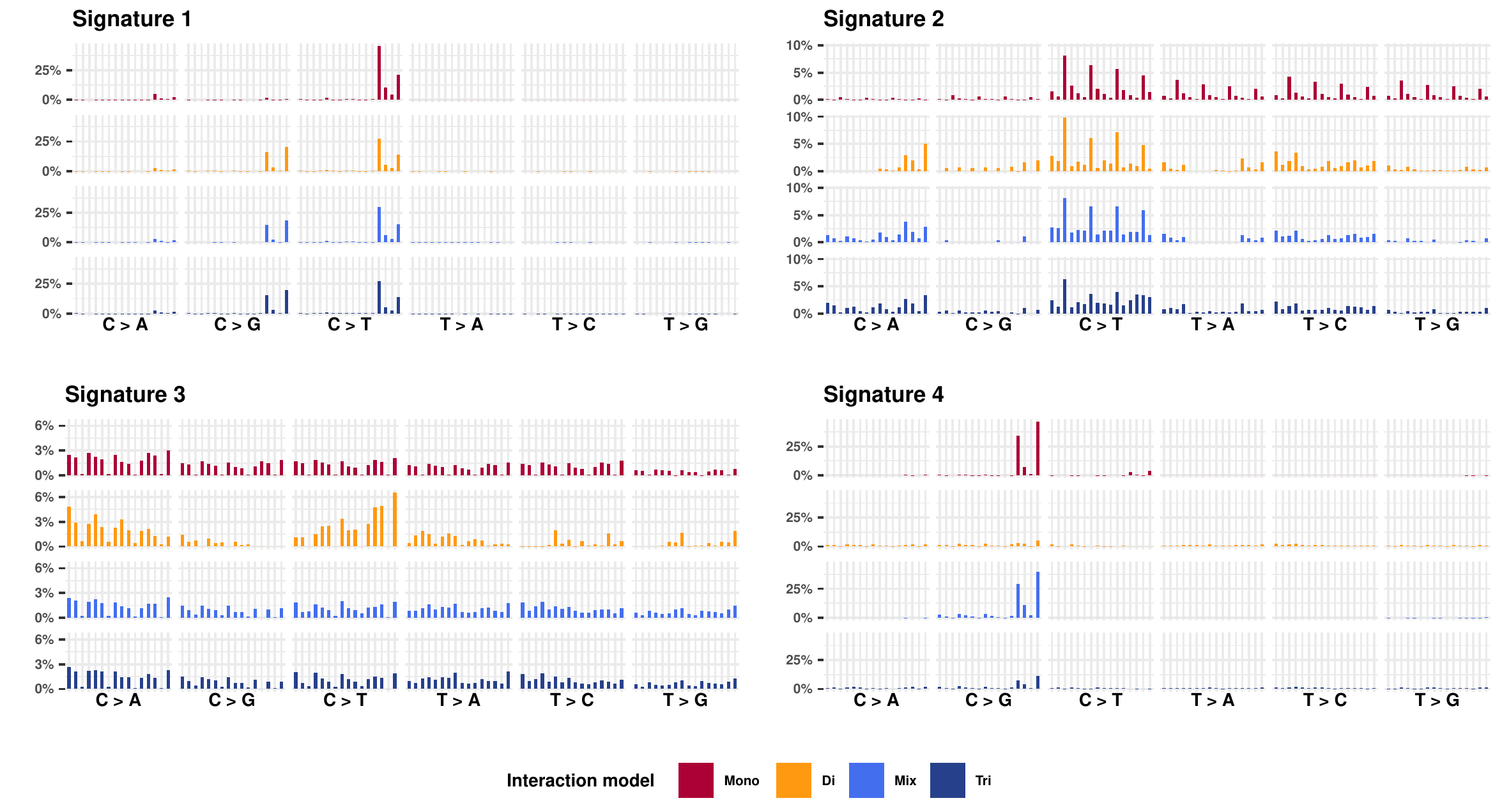} \\
    \noindent {\bf B. Parametric bootstrap} \vspace{1mm} \\
    \includegraphics[width = 0.9\textwidth]{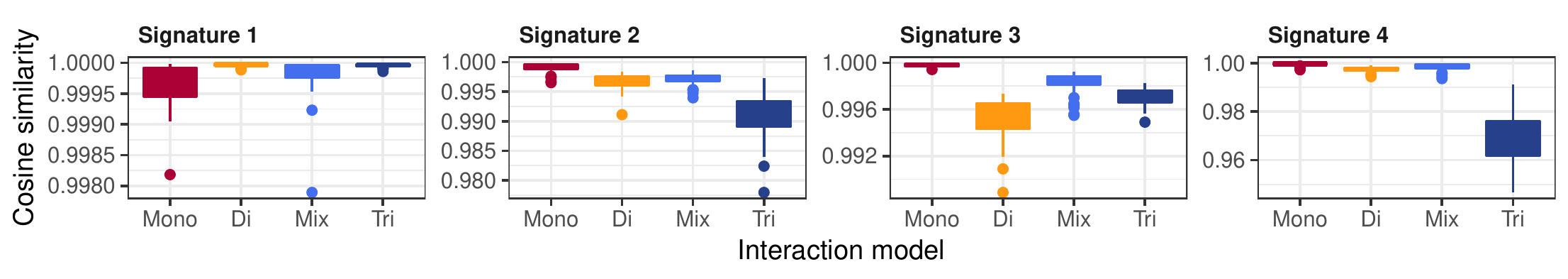}
 \end{tabular}
    \caption{\textbf{A.} Comparison of the four signatures for the four highlighted models in Figure~\ref{fig:BRCA21models}. The four models are three models where all four signatures are mono-nucleotide, di-nucleotide or tri-nucleotide, and one mixture model where one signature is tri-nucleotide, two are di-nucleotide and one is mon-nucleotide. \textbf{B.} The cosine similarity for reconstructing the signatures with parametric bootstrapping. }
    \label{fig:BRCA21signatures}
\end{figure}

In Figure~\ref{fig:BRCA21signatures}A, we show the four signatures for the four different models and  
in Figure~\ref{fig:BRCA21signatures}B we show the cosine similarity for reconstructing the signatures from the parametric bootstrap procedure.
Signatures~1, 2 and 4 are very similar for the di-nucleotide model and the tri-nucleotide model. 
These three signatures also give very high cosine similarity for reconstructing the di-nucleotide model, which means they are very stable. The mono-nucleotide model have very stable signatures as the cosine similarity is high, but the signatures are also very different from the signatures in the full model. For the mono-nucleotide signature 2 it is clear to observe that the few parameters in the model are restricting signature to have equal peaks within each base mutation. The only difference between the values for all the $C>T$ and $T>A$ mutation types is a multiplicative constant. This is also evident from the lack of fit the mono-nucleotide model has to the data. The signatures in the mix model are all very similar to the full model except minor differences, and they are also more stable than the signatures in the full model. 

\begin{figure}[htb!]
    \centering
    \includegraphics[width = 0.85\textwidth]{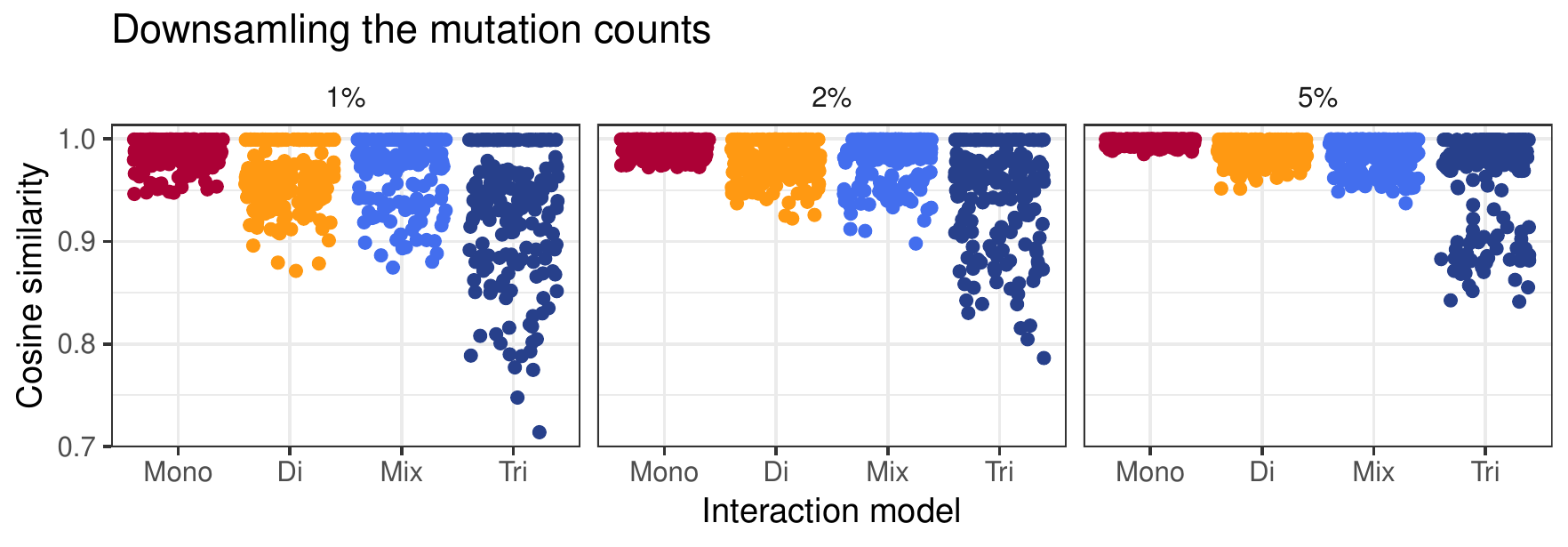}
    \caption{The cosine similarity between the recovered exposures from down-sampled BRCA21 data compared to the exposures from the original BRCA21 data.}
    \label{DownSamplingFig}
\end{figure}

Finally, we also investigate the stability of the exposures from the different parametrizations of the signatures. We again compare the four different models from Table~\ref{BRCA21tbl}. We fix the four signatures to the values obtained from the full data and down-sample to 1~percent, 2~percent or 5~percent of the total original mutation counts. In each experiment we then re-estimate the exposures for the four signatures of the four interaction models by minimizing the generalized Kullback-Leibler divergence. In Figure~\ref{DownSamplingFig} we show the cosine similarity between the original and re-estimated exposures from the down-sampled data for the four different models. We observe that the exposures for the di-nucleotide model are better recovered than the exposures for the tri-nucleotide model. In general, we observe that a simpler parametrization gives a more robust estimation of the exposures. This feature could be important if the exposures are used for deciding upon diagnosis or treatment of cancer patients. 
\subsection{Analysis of BRCA214 data} \label{sec:brca214}
\begin{figure}[h!]
    \centering
    \includegraphics[width = 0.57\textwidth]{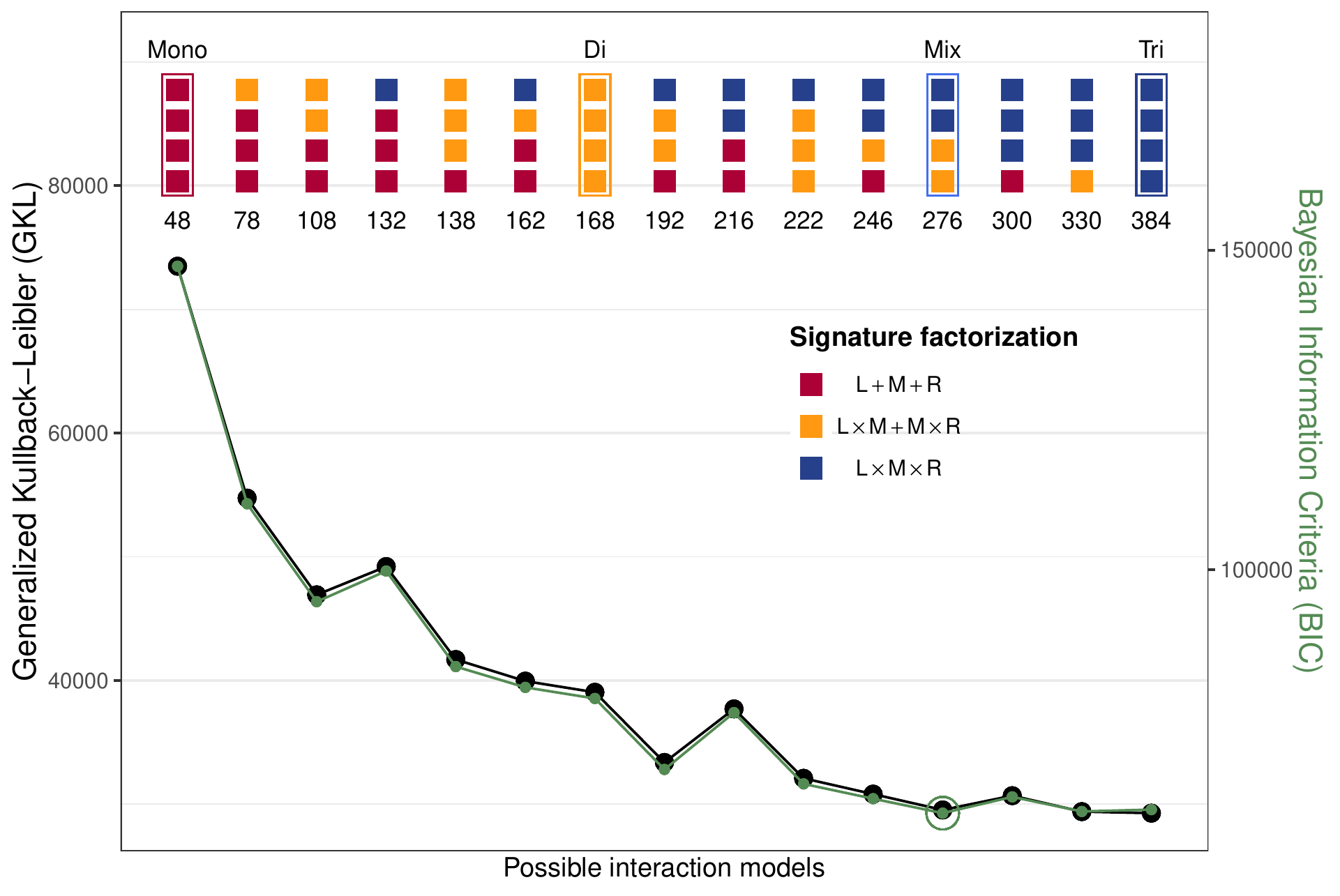} 
    \caption{Fit to 214 breast cancer patients for all possible interaction models. The Generalized Kullback-Leibler (GKL) and Bayesian Information Criteria (BIC) for all 15 models with 4 signatures. The models are ordered according to the total number of parameters for the 4 signatures. The model with the smallest BIC is indicated, and consists of two di-nucleotide signatures and two tri-nucleotide signature.}
    \label{fig:GKLBRCA214}
\end{figure}

The BRCA214 data set consists of $T=96$ mutation types and $N=214$ breast cancer patients, which means 193 additional patients compared to BRCA21. The values of the GKL for the various interaction models in Figure~\ref{fig:GKLBRCA214} show that the pure mono-nucleotide signatures model (Mono in Figure~\ref{fig:GKLBRCA214}) is a poor fit to the data. As for BRCA21 a good fit is obtained with one mono-nucleotide signature, two di-nucleotide signatures and one tri-nucleotide signature, but the smallest BIC is obtained for a model with two di-nucleotide signatures and two tri-nucleotide signatures.

\begin{figure}[h!]
    \centering  
    \begin{tabular}{l}
    \noindent {\bf A. Inferred signatures for the BRCA214 data set} \vspace{1mm} \\
    \includegraphics[width = 0.84\textwidth]{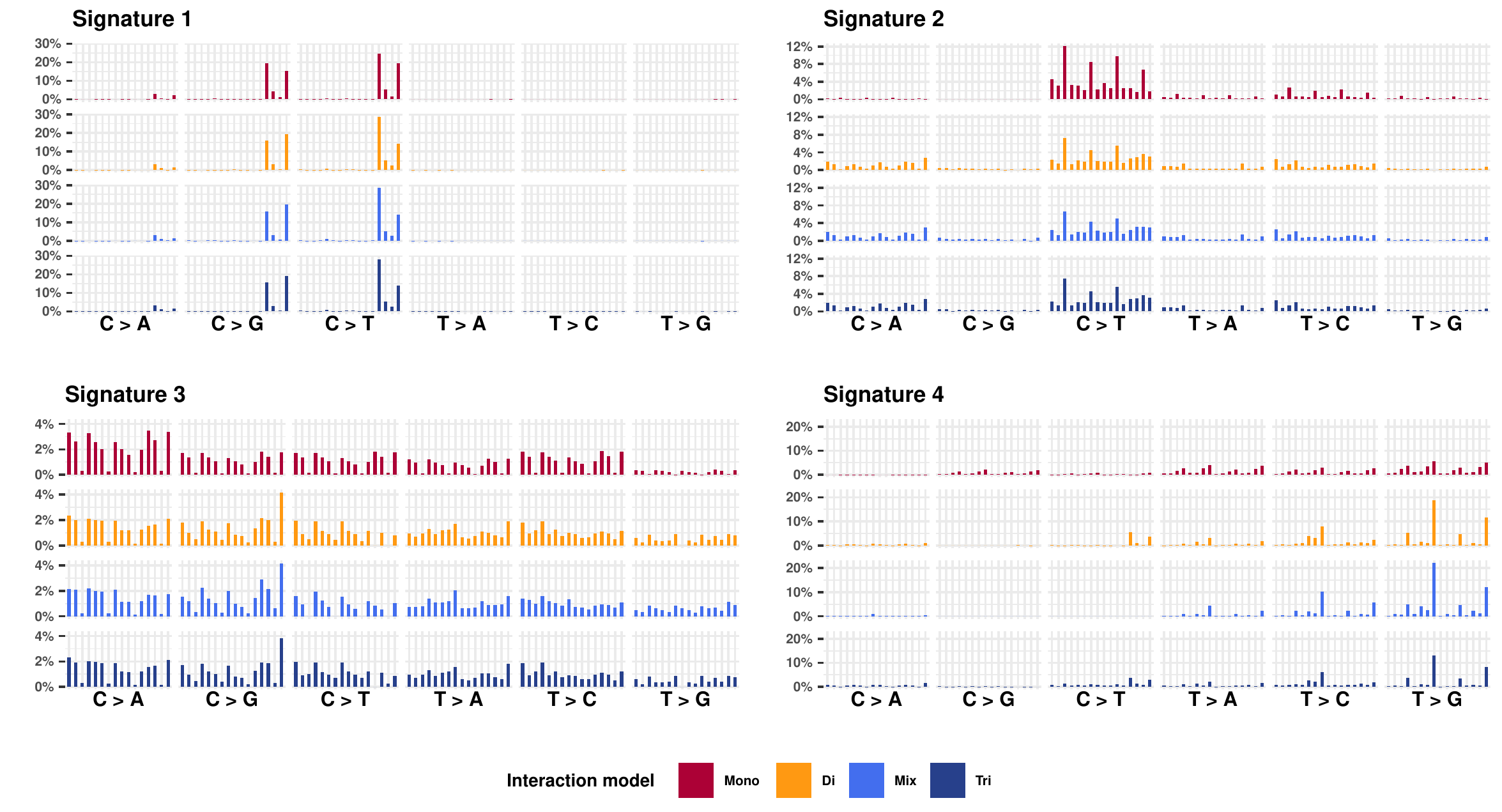} \\
    \noindent {\bf B. Parametric bootstrap} \vspace{1mm} \\
    \includegraphics[width = 0.84\textwidth]{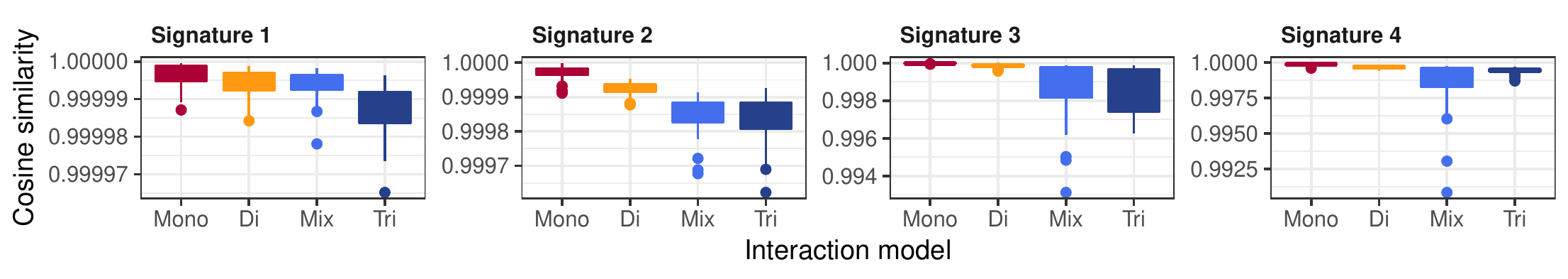}
    \end{tabular}
    \caption{\textbf{A.} Comparison of the four signatures for the four models that are highlighted in Figure~\ref{fig:GKLBRCA214} as Mono, Di, Mix and Tri. \textbf{B.} The cosine similarity for reconstructing the signatures with parametric bootstrap.}
    \label{fig:SignaturesBRCA214}
\end{figure}

When we compare the signatures in Figure~\ref{fig:BRCA21signatures} for the 21 patients with the signatures for the 214 patients in Figure~\ref{fig:SignaturesBRCA214}, we see that the added patients mainly have changed Signature~4, whereas the remaining three signatures all look fairly similar. The average cosine similarities for the four signatures are respectively $0.95$, $0.93$, $0.86$ and $0.17$. Many of the signatures are therefore recovered even for the small number of patients, but the increase in patients makes them more robust. 

We also carried out parametric bootstrap for the BRCA214 data set and the result is shown in Figure~\ref{fig:SignaturesBRCA214}. 
Now that the mixture model with the smallest BIC consists of two di-nucleotide signatures and two tri-nuclotide signatures, which is close to the full model, we observe in Figure~\ref{fig:SignaturesBRCA214} that the signatures for the mixture model are more unstable. The variability in cosine similarity for reconstructing the mutational signatures for the mono-nucleotide and di-nucleotide models is now always smaller than for the mixture and tri-nucleotide models.   


\newpage
\subsection{Analysis of UCUT26 data} \label{sec:ucut}
The UCUT26 data contains information about the two flanking bases at each side.
The UCUT26 count matrix has $T=6\cdot4^4=1536$ mutation types and $N=26$ patients. The data consists of 14.715 somatic mutations, and the number of non-zero entries in the count matrix is $n_{\rm obs}=5260$. 
\begin{table}[h!]
\centering
\resizebox{\textwidth}{!}{%
\begin{tabular}{crrrr}
  \hline
  & Number of & Model & & Model \\
  & parameters & complexity & Fit to data & selection \\
 Model for the two signatures & $n_{\rm prm}$ & $n_{\rm prm}\log n_{\rm obs}$ & ${\rm GKL}$  & ${\triangle \rm BIC}$ \\ \hline \hline
 $L_2+L_1+M+R_1+R_2$ & $2 \cdot 18=36$ & 308 & 10422 & 2116 \\ 
 $L_1\times M \times R_1$ & $2 \cdot 96=192$ & 1645 & 10182 & 2972 \\\hline
 $L_2+L_1\times M+M \times R_1+R_2$ & $2 \cdot 48=96$ & 823 & 9788 & 1363 \\
 $L_2+L_1\times M \times R_1+R_2$ & $2 \cdot 102=204$ & 1748 & 9438 & 1588 \\ 
 $L_2\times L_1+L_1\times M+M\times R_1+R_1\times R_2$ $(a)$ & $2 \cdot 66=132$ & 1131 & 9008 & 111 \\ 
 $L_2\times L_1+L_1\times M \times R_1+R_1\times R_2$ $(b)$ & $2 \cdot 120=240$ & 2056 & 8658 & 336 
 \\ \hline
 $L_2\times L_1\times M \times R_1\times R_2$ & $2 \cdot 1536=3072$ & 26321 & 6982 & 21249 
 \\ \hline \hline
  Mixture of signature $(a)$ and $(b)$ & $120 + 66 = 186$ & 1594 & 8721 & 0
\end{tabular}
}
\caption{Summary statistics for the seven basic models for the UCUT26 data where both signatures have the same parametrization. The models are ordered according to their GKL value. The number of signatures is $K=2$ and the number of observations is $n_{\rm obs}=5260$. At last the mixture model with the smallest BIC is also depict, which all the other BIC values are compared to. }
\label{UCUT26tbl}
\end{table}

The penta-nucleotide interaction signature $L_2\times L_1\times M \times R_1\times R_2$ has 1536 parameters (recall Table~\ref{tab:ModelTwo}), and this many parameters inevitably results in over-fitting for the UCUT26 data set. This model is included as a control to show that the full parametrization gives an extremely high BIC value compared to the other models. A parametrization with much fewer parameters is needed for inferring robust signatures, and the mono-nucleotide interaction signatures $L_2+L_1+M+R_1+R_2$ from \cite{shiraishi2015} was originally developed for this purpose. Here, we also consider a di-nucleotide signature of the type $L_2\times L_1+L_1\times M+M\times R_1+R_1\times R_2$, and three signatures that have a combination of interaction terms $L_2+L_1\times M+M \times R_1+R_2$, $L_2+L_1\times M \times R_1+R_2$ and $L_2\times L_1+L_1\times M \times R_1+R_1\times R_2$. Finally, we include the tri-nucleotide signature $L_1\times M \times R_1$ to investigate whether the two immediate flanking nucleotides are sufficient for explaining the probability of a somatic cancer mutation.  

The seven models where both signatures have the same parametrization and their performance statistics when applied to the UCUT26 data set are summarized in Table~\ref{UCUT26tbl}. 
The table summarizes the number of parameters $n_{\rm prm}$, model complexity $n_{\rm prm}\log n_{\rm obs}$, model fit GKL, and the differences between the model selection measure BIC and the smallest obtained BIC. We observe that two immediate flanking nucleotides (one at each side) are not sufficient for explaining the mutation patterns: the $L_1\times M \times R_1$ model has the same poor fit to data as the mono-nucleotide model despite having more than five times as many parameters. The four models  $L_2+L_1\times M+M \times R_1+R_2$, $L_2+L_1\times M \times R_1+R_2$, $L_2\times L_1+L_1\times M+M\times R_1+R_1\times R_2$ and $L_2\times L_1+L_1\times M \times R_1+R_1\times R_2$ all show a relatively good fit to the data, but the $L_2+L_1\times M \times R_1+R_2$ model is penalized for the many parameters. Finally, the $L_2\times L_1+L_1\times M+M\times R_1+R_1\times R_2$ and $L_2\times L_1+L_1\times M \times R_1+R_1\times R_2$ model have a superior fit to the data compared to the other models, and does not contain too many parameters. We note that these two models are the only models with di-nucleotide interaction between the two left flanking nucleotides (both models contain the term $L_2\times L_1$) and the two right flanking nucleotides (the term $R_1\times R_2$), and conclude that these interaction terms are important for quantifying the probability of a somatic mutation in this cancer type.

\begin{figure}[t!]
    \centering
    \includegraphics[height = 0.27\textheight]{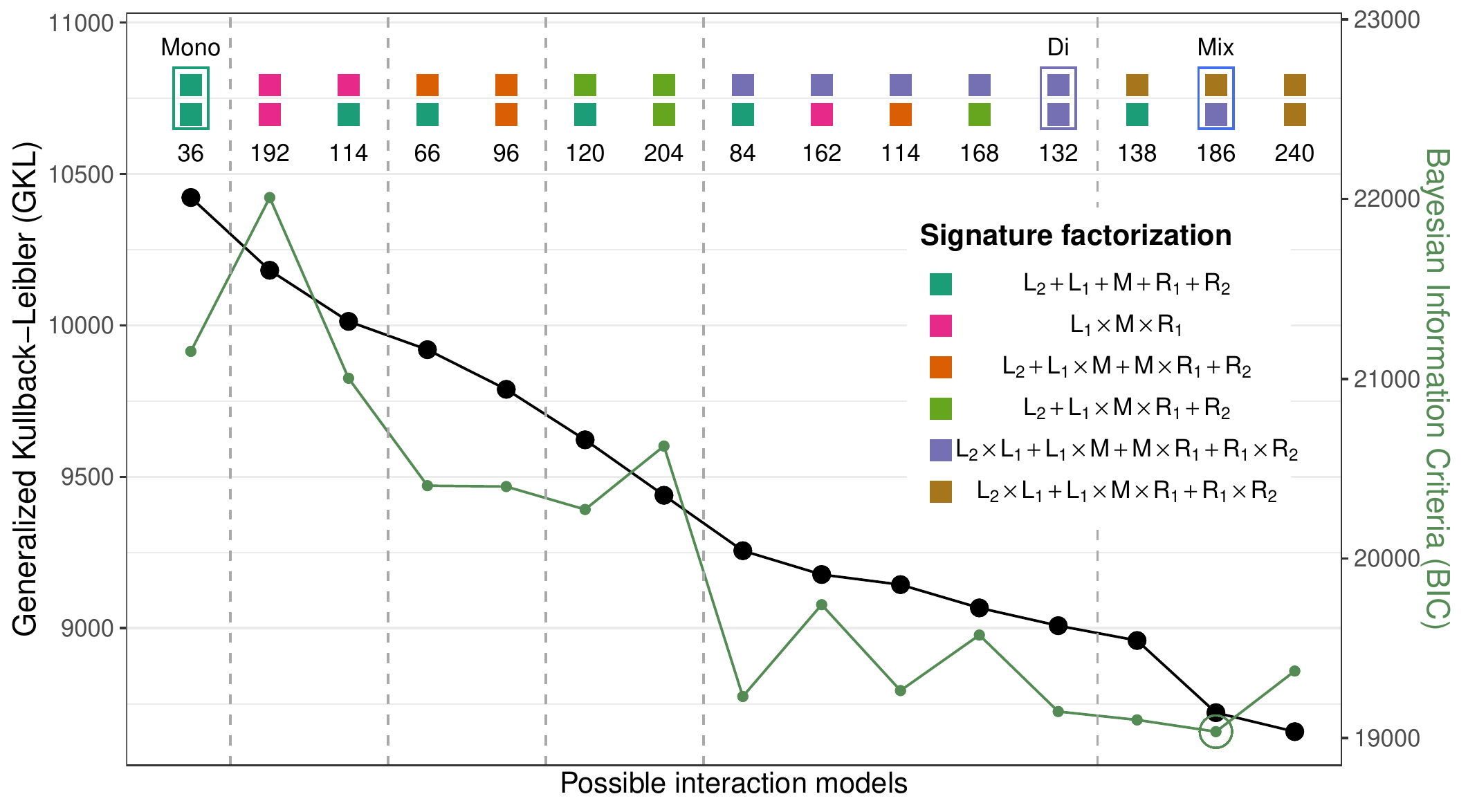}
    \caption{The Generalized Kullback-Leibler for all 21 models with two signatures for the UCUT26 data set. The models are ordered according to GKL values, which also orders the models by the first signature.}
    \label{fig:UCUTmodels}
\end{figure}

We also consider parametrizations of the signature matrix where the two signatures have different parametrizations. The GKL and BIC for 15 different combinations of the seven parmetrizations is summarized in Figure~\ref{fig:UCUTmodels}. Here, we have ordered the models by the GKL value as this automatically groups the different signature parametrizations. We have left out the penta-nucleotide signature, as it gives extremely high BIC values due to the many parameters in the model. Similar to our finding for the BRCA21 data set, we observe that two mono-nucleotide signatures $L_2+L_1+M+R_1+R_2$ give a poor fit to the data. We emphasize that two tri-nucleotide signatures $L_1\times M \times R_1$ or a mixture of the two all have a poor fit to the data, which means the information about the flanking nucleotides two positions away from the mutation is important. We find that a mixture between the two parametrizations $L_2\times L_1 +L_1\times M+M\times R_1+R_1\times R_2$ and $L_2\times L_1 +L_1\times M \times R_1+R_1\times R_2$ fits the data very well despite the rather few parameters; this mixture model has the smallest BIC value. 
\begin{figure}[t!]
    \centering
    \includegraphics[width = 0.75\textwidth]{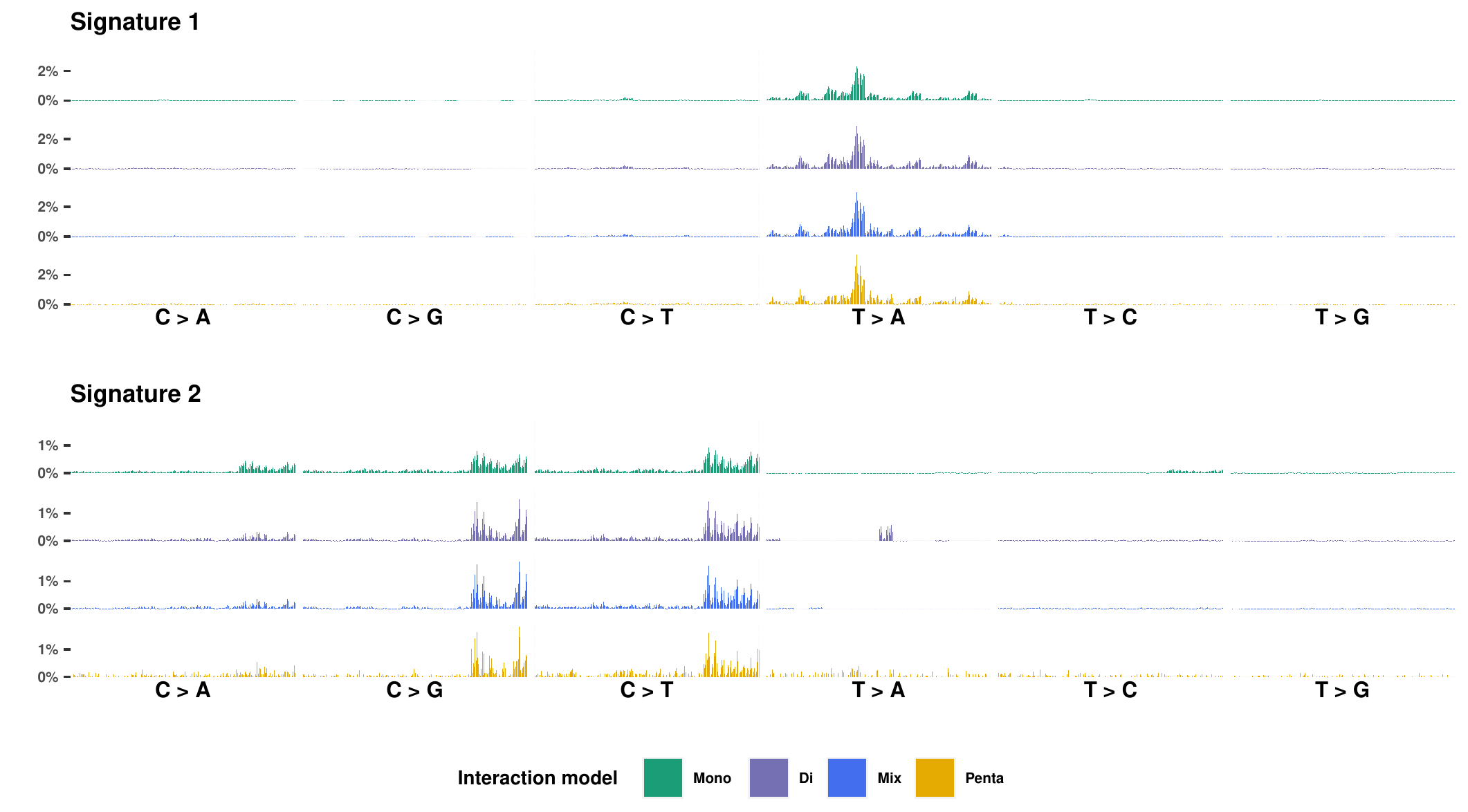}
    \raisebox{0.28\height}{ \includegraphics[width = 0.2\textwidth]{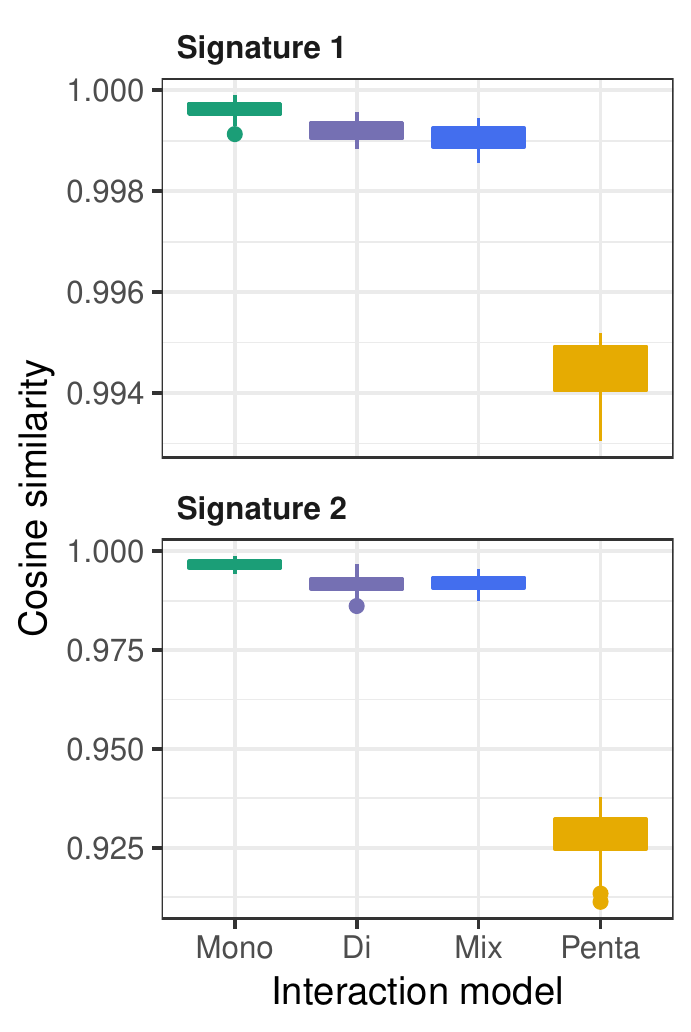}}
    \caption{The Generalized Kullback-Leibler and Bayesian Information Criteria for 15 models with two signatures for the UCUT26 data set. The models are ordered according to GKL values, which also orders the models by the first signature.}
    \label{fig:UCUTsig}
\end{figure}

In Figure~\ref{fig:UCUTsig} the two signatures are visualized for the mono, di, mix and penta model. For the mixture model, signature 1 is described by the di- and tri-nucleotide interactions and signature 2 only by the di-nucleotide interactions. In the original study in \cite{Hoang2013} they identify signature 1 as a novel mutation signature that predominantly contains $T > A$ substitutions at CpTpG sites caused by aristolocthic acids. This dependence of a single tri-nucleotide interaction is potentially why the signature is best parametrized when including tri-nucleotide interactions. 

The cosine similarities for reconstructing the signatures from parametric bootstrap show that the penta-nucleotide signatures are much worse at reconstructing the same signatures. Again, this indicates the problem with too many parameters in the model. On the other hand, the model with two di-nucleotide signatures and the mixture model are almost just as stable as the mono-nucleotide signatures, but gives a much better fit to data. 

These findings demonstrate the relevance of our flexible framework for mutational signatures. The di--nucleotide signatures provide a better fit to the data and are biologically more plausible than mono-nucleotide signatures, and the parametrization is more stable than the parameter-rich signatures with interaction terms higher than or equal to three. The ability to allow a combination of signatures is also advantageous.
\section{Methods} \label{sec:Methods}
Recall the NMF set-up from Section~\ref{sec:Param} where the mutation counts in the data matrix $V \in \mathbb{N}_+^{N \times T}$ are modelled by the Poisson distribution 
$V_{nt} \sim \text{Pois}((WH)_{nt})$. The data log-likelihood in \eqref{eq:datalik} is analytically intractable, but we can view the problem as a missing data problem where the missing information is the assignment of each mutation to a signature. If this information was available, then a likelihood analysis would be easy, and therefore the EM-algorithm (\cite{DempsterLairdRubin1977}) applies. In this section we describe the EM-algorithm for estimating the parameters in non-negative matrix factorization. We first describe the EM-algorithm for the traditional model where the only constraints on the exposure matrix $W$ and signature matrix $H$ in the matrix factorization are that the entries must be non-negative (e.g. \cite{cemgil2008bayesian}). Second, we extend the EM-algorithm to the situation where the signatures are parametrized according to \eqref{logH}.   
\subsection{EM-algorithm for traditional non-negative matrix factorization}
Given a data matrix $V \in \mathbb{N}_+^{N \times T}$ the aim of NMF is to find a non-negative factorization $WH$, where $W \in \mathbb{R}_+^{N \times K}$ and $H \in \mathbb{R}_+^{K \times T}$ approximates of our data $V$ i.e. $V \approx WH$. 
The rank $K$ of the factorization is often chosen magnitudes smaller than the minimum of $N$ and $T$. A larger $K$ obviously gives a better fit, but would potentially overfit the data. In traditional NMF all the entries in $W$ and $H$ are free parameters that need to be estimated. Later we will reduce the number of free parameters in $H$, but first we describe the traditional estimation of $W$ and $H$.

A challenge with the likelihood function in \eqref{eq:datalik} is that it is only convex in either $W$ or $H$, but not in both matrices together. This means we cannot find a closed form solution for the maximum likelihood estimates of $W$ and $H$, and instead we use the EM-algorithm. 
For the EM-algorithm we introduce the latent variables 
\[Z_{nkt} \sim \text{Pois}(W_{nk}H_{kt})  \]
which is the mutational count from each of the $K$ signatures for each observation, such that the total number of mutations for a cancer patient $n$ of a certain type $t$ is given by
\[V_{nt} = \sum_{k=1}^K Z_{nkt} \sim \text{Pois}((WH)_{nt}).\] 
The complete log-likelihood is given by
\begin{align}
 \ell(W,H;Z) &= \sum_{n=1}^N \sum_{t=1}^T \sum_{k=1}^K \left\{ Z_{nkt} \log(W_{nk} H_{kt}) - W_{nk} H_{kt} - \log(Z_{nkt}!) \right\}   \label{eq:comlik} \\ 
& \equiv \sum_{k=1}^K \sum_{t=1}^T \left( \sum_{n=1}^N Z_{nkt} \right) \log(H_{kt})  
 + \sum_{k=1}^K \sum_{n=1}^N \left\{ \left( \sum_{t=1}^T Z_{nkt} \right) \log(W_{nk}) - W_{nk} \right\} 
\end{align}
where we use that signatures are probability distributions that sum to one, $\sum_{t=1}^T H_{kt} = 1$, and $\equiv$ means that the statement is true up to the additive constant $\sum_{n=1}^N \sum_{t=1}^T \sum_{k=1}^K \log(Z_{nkm}!)$. \\
\textbf{E-step:} 
For fixed values $W^i$ and $H^i$ this step finds the expected value of the latent variables $\{Z_{nkt}\}$ conditional on the data $V$. The distribution of $\{Z_{nkt}\}$ conditional on their sum is given by the multinomial distribution  
\begin{equation*}
    (Z_{n1t}, \dots , Z_{nKt}) \bigg| V_{nt} = { \small \sum_{k=1}^K Z_{nkt} } \sim \text{Multi}\left(V_{nt}, \frac{1}{(WH)_{nt}}\left(W_{n1}H_{1t}, \dots , W_{nK}H_{Kt} \right) \right),
\end{equation*}
which implies that
\[
\mathbb{E}_{W^i,H^i}[Z_{nkt}|V] = 
\mathbb{E}_{W^i,H^i}[Z_{nkt}|V_{nt}] = 
V_{nt} \frac{W^i_{nk}H^i_{kt}}{(W^i H^i)_{nt}}.
\]
Replacing $\{Z_{nkt}\}$ with their expected values $\mathbb{E}_{W^i,H^i}[Z_{nkt}|V]$ gives the expected complete log-likelihood  
\begin{align}
 Q(W,H|W^i,H^i) &= \sum_{k=1}^K \sum_{t=1}^T \left( \sum_{n=1}^N \mathbb{E}_{W^i,H^i}[Z_{nkt}|V] \right) \log(H_{kt}) \label{eq:Hlik} \\
 & \quad + \sum_{k=1}^K \sum_{n=1}^N \left\{ \left( \sum_{t=1}^T \mathbb{E}_{W^i,H^i}[Z_{nkt}|V] \right) \log(W_{nk}) - W_{nk} \right\}  \label{eq:Wlik}
\end{align}
\textbf{M-step:} 
The first term of the expected complete log-likelihood \eqref{eq:Hlik} is recognised as $K$ independent multinomial log-likelihood functions and the second term \eqref{eq:Wlik} is recognised as $N \cdot K$ Poisson log-likelihoods. Maximum of the expected complete log-likelihood with respect to $W$ and $H$ is therefore given by
\begin{align}
  H^{i+1}_{kt} = 
  \frac{ \sum_{n=1}^N 
  \mathbb{E}_{W^i,H^i}[Z_{nkt}|V] }{ \sum_{t=1}^T \sum_{n' = 1}^N \mathbb{E}_{W^i,H^i}[Z_{n'kt}|V] } = 
  \frac{ \sum_{n=1}^N V_{nt} \frac{W^i_{nk}H^i_{kt}}{(W^iH^i)_{nt}} }{\sum_{t = 1}^T \sum_{n'=1}^N V_{n't} \frac{W^i_{n'k}H^i_{kt}}{(W^iH^i)_{n't}}}  
  \label{Hupdate}
\end{align}
and
\begin{align}
  W_{nk}^{i+1} = 
  \sum_{t=1}^T \mathbb{E}_{W^i,H^i}[Z_{nkt}|V] =
  \sum_{t=1}^T V_{nt} \frac{W^i_{nk}H^i_{kt}}{(W^iH^i)_{nt}}.
\end{align}
The expected value of $\{Z_{nkt}\}$ from the E-step is also inserted, which means these updates include both steps of the EM-algorithm to find the optimal estimates $W$ and $H$. The entire EM-algorithm with initialization and stopping criteria to obtain the optimal parameters is summarized in Algorithm~\ref{alg:EM}. The updates are written in vector form for $H$ and matrix form for $W$. Note that $\otimes$ and division means entry wise multiplication and division, the vector $\textbf{1}$ is of length $T$ and consists only of ones, $W_k$ is the $k$'th column of $W$, and $H_k$ is the $k$'th row of $H$. We stop the EM-algorithm when the data log-likelihood after a full update of $W$ and $H$ is smaller than a threshold $\epsilon$.

\begin{algorithm}
\SetAlgoLined
 Given data matrix $V$, rank $K$ and threshold $\epsilon$. \\
 Initialize $W^1$ and $H^1$ with random entries. \\
 \For{ $i = 1,2,3, \dots $ }{
 \For{ $k = 1,\dots, K$}{
 Update each signature
 \begin{eqnarray}
  H_k^{i+1} &= \frac{H_k^i \otimes \left( (W_k^i)' \frac{V}{W^iH^i} \right)}{ \textbf{1}' \left( H_k^i \otimes \left( (W_k^i)' \frac{V}{W^iH^i} \right) \right) } \label{eq:Hupdate} 
  \end{eqnarray} 
  }
  Update exposures
  \begin{eqnarray*}
    W^{i+1} &= W^i \otimes \left( \frac{V}{W^iH^i} (H^i)' \right)
  \end{eqnarray*}
   \textbf{stop if} $\frac{\ell(W^{i+1},H^{i+1};Z)-\ell(W^i,H^i;Z)}{\ell(W^{i+1},H^{i+1};Z)} < \epsilon$
 }
\caption{General EM-algorithm to estimate exposures $W$ and signatures $H$.}
\label{alg:EM}
\end{algorithm}
\subsection{EM-algorithm for parametric non-negative matrix factorization}
Another parametrization of the signatures $H_1, \dots H_K$ requires a change in update \eqref{Hupdate} which was based on maximizing \eqref{eq:Hlik}. The parametrization of the signatures are given by the design matrices $X_1, \dots X_K$. 
Recall that the number of mutations from a specific signature for each observation is given by the latent variables $\{Z_{nkt}\}$. 
We observe that we again have $K$ independent multinomial log-likelihood terms that we can maximize separately. 
Define
\[Y^i_{kt} = \sum_{n=1}^N \mathbb{E}_{W^i,H^i}[Z_{nkt}|V], \]
which is the expected number of mutations at the $i$'th iteration for signature $k$ of type $t$. We now suppress the superscript $i$ and subscript $k$ by introducing the simple notation $y_t=Y_{kt}^i$ and $h_t = H_{kt}$.
In parallel to \eqref{eq:Hlik} we need to maximize 
\begin{eqnarray*}
  \sum_{t=1}^T y_t \log(h_t)
\end{eqnarray*}
with respect to $\beta$ where we set 
\begin{eqnarray}
  h_t=\frac{\exp((X\beta)_t)}{\sum_{t=1}^T \exp((X\beta)_t)},
  \label{eq:multireg}
\end{eqnarray}
and again we have suppressed the dependency on $k$ in both $X$ and $\beta$. Instead of estimating $\beta$ in this model, we use the 'Poisson Trick' (see e.g. \cite{Lee2017} or Section 6.4 in~\cite{McCullaghNelder1989}).
The 'Poisson Trick' means that the log-linear Poisson model 
\begin{equation}
   \log (y_t) = (X \beta)_t, \quad t=1, \dots , T,
   \label{eq:poisreg}
\end{equation}
is equivalent to the multinomial response model with probabilities given by \eqref{eq:multireg}. We therefore determine the maximum likelihood estimate of $\beta$ by fitting the log-linear Poisson model instead of the multinomial response model.
The full EM-algorithm is presented in matrix form in Algorithm \ref{alg:EMglm}. 

\begin{algorithm}[ht] 
\SetAlgoLined
Given data matrix $V$, rank $K$, design matrices $X_1, \dots , X_K$, and threshold $\epsilon$.\\
Initialize $W^1$ and $H^1$ with random entries. \\
 \For{ $i = 1,2,3, \dots $ }{
 \For{ $k = 1,\dots, K$}{
 Update each signature
       $$\bm{y}_k^{i} = H_k^i \otimes \left( (W_k^i)' \frac{V}{W^i H^i} \right) $$
      Fit the log-linear Poisson regression \\
      \begin{equation}
    \log ( \bm{y}_k^{i}) = X_k \beta_k^i
\end{equation}
      for estimating $\beta_k^i$ and set 
     $$H_k^{i+1} = \frac{ \exp(X_k \hat{\beta}_k^i) }{ \textbf{1}' \exp(X_k \hat{\beta}_k^i) } $$\\
 }
 Update exposures
\begin{equation*}
     W^{i+1} = W^i \otimes \left( \frac{V}{W^i H^i} (H^i)' \right)
\end{equation*}
\textbf{stop if} $\frac{\ell(W^{i+1},H^{i+1};Z)-\ell(W^i,H^i;Z)}{\ell(W^{i+1},H^{i+1};Z)} < \epsilon$
 }
\caption{Parametric EM-algorithm to estimate exposures $W$ and signatures $H$.}
\label{alg:EMglm}
\end{algorithm}

Estimation of $\beta$ in \eqref{eq:poisreg} is obtained by fitting the log-linear Poisson model using the Newton-Raphson method, and for clarity we provide the details. The log-likelihood function for the Poisson model with design matrix $X$ of dimension $T\times S$, parameter vector $\beta$ of length $S$ and data vector $y=(y_1,\ldots,y_T)$ of length $T$ is given by
\[
    \ell (\beta;y,X) \equiv 
    \sum_{t=1}^T  \big\{ y_t (X\beta)_t-\exp \left( (X\beta)_t \right) \big\}.
\]
A closed form solution for the maximum likelihood estimate is in general not available, but we can use the Newton-Raphson method.
The gradient and the Hessian of the log-likelihood function are  
\begin{align*}
   \frac{\partial \ell}{ \partial \beta} = X' \left\{ y-\exp(X \beta) \right\} 
   \quad {\rm and} \quad
   \frac{\partial^2 \ell}{ \partial \beta' \partial \beta} = - X' A X,
\end{align*}
where $A=A(\beta)$ is a diagonal matrix of dimension $S\times S$ with $\exp(\sum_{t=1}^T \beta_s X_{st})$, $s=1,\ldots,S,$ on the diagonal. 
The Newton-Raphson update is given by
\begin{equation*}
  \beta^{i+1}=
  \beta^i+(X' A^i X)^{-1} X' \big\{ y-\exp(X \beta^i) \big\},
\end{equation*}
where $A^i=A(\beta^{i})$, which can be re-written as
\begin{align*}
  \beta^{i+1} &= 
  (X' A^i X)^{-1} X' A^i 
  \big[ X\beta^i + (A^{i})^{-1}\{y - \exp(X \beta^i) \}\big] \\
  &= (X' A^i X)^{-1} X' A^i \upsilon^i,
\end{align*}
where
\begin{equation*}
    \upsilon^i = X \beta^i + (A^i)^{-1} \left\{ y- \exp(X \beta^i) \right\}.
\end{equation*}
This means that the update is the solution to the weighted least square problem
\begin{eqnarray*}
  \beta^{i+1}=\arg \min_{\beta} ||(A^i)^{1/2}(\upsilon-X\beta^i)||^2.
\end{eqnarray*}
In our implementation in R we call the built-in method to solve the weighted least squares problem. 

To accelerate the EM-algorithm we have both made a version that uses the R package SQUAREM \citep*{Du2020} and another version implemented in C++. To escape local maxima of the likelihood function we typically start the algorithm 100 or even 500 times and run each of them for 100 or 500 iterations before we identify a minimum. We then let the identified minimum iterate until convergence.
\section{Discussion}
We have presented new flexible and biologically plausible parametrizations of mutational signatures. The di-nucleotide interaction signatures strike a good balance between a satisfactory fit to the data and statistically stable and simple signatures. We have emphasized this through the analysis of the three data sets BRCA21, BRCA214 and UCUT26. Besides making it possible to parametrize the mutational signatures, we also mix the different parametrizations for the signature matrix. This makes the parametrization even more flexible as we allow each signature to have its own parametrization. We believe this could be an important step forward in order to be able to specify and understand the most important interaction terms between the mutation and the flanking nucleotides.     

We have considered a number of signatures $K = 2$ for the UCUT data set and $K=4$ for the BRCA data sets. For higher values of $K$ a full investigation of all models is computationally time consuming. 
A natural strategy would be to use a factor diagram as in Figure~\ref{fig:factordiagram} and start from the full parametrization and then test if the signature parmetrizations can be simplified. This strategy is similar to a backward stepwise selection procedure, while the proposed procedure correspond to a best subset selection procedure. This method could be even more relevant if we would like to identify specific important interactions like the CpTpG tri-nucleotide interaction for the UCUT data. Another thing to note is that the few number of signatures could have an influence on the parametrization. Having several mutational processes included in the same mutational signature could explain the need for more complex parametrizations in some cases.  

Our main goal has been statistical robustness and interpretation of the signatures, and this is achieved by biologically plausible constraints on the parameters: we allow each signature to contain mono-, di- or tri-nucleotide interaction terms. An alternative to the supervised constraints would be a more unsupervised strategy for learning signatures. For example, one could impose sparseness on the signatures in the spirit of \cite{Lal2020}.  

We have focused on finding a single parametrization for each signature where interpretation is easy. This is useful when the aim is to recover the true underlying biological mechanisms that cause the various signatures (e.g. UV-light or tobacco smoking).  
Model averaging over different parametrizations would make sense if the goal is a statistically robust signature where interpretation is less important (e.g. classification of a genomic region based on the mutation profiles). The BIC values are rather similar for many of the models, suggesting that model averaging could be useful.
  
Our flexible framework also allows inclusion of other factors known to have an impact on somatic mutations such as replication timing \citep{WooLi2012}, expression level \citep{lawrence2013} or general conservation of the position when compared to other species \citep{Bertl2018}. This was not included in this paper, but would be fairly simple to add into the model.   
\section*{Acknowledgements}
We thank Camilla Provstgaard Kudahl and Maiken Bak Poulsen for valuable initial results and discussions. We are grateful to Marta Pelizzola and Gustav Alexander Poulsgaard for very helpful comments on the manuscript.
\bibliographystyle{apalike}
\bibliography{ref}

\begin{thebibliography}{}

\bibitem[Alexandrov et~al., 2016]{Alexandrov2016}
Alexandrov, L.~B., Ju, Y.~S., Haase, K., Van~Loo, P., Martincorena, I.,
  Nik-Zainal, S., Totoki, Y., Fujimoto, A., Nakagawa, H., Shibata, T.,
  Campbell, P.~J., Vineis, P., Phillips, D.~H., and Stratton, M.~R. (2016).
\newblock Mutational signatures associated with tobacco smoking in human
  cancer.
\newblock {\em Science}, 354(6312):618--622.

\bibitem[Alexandrov et~al., 2020]{alexandrov2020}
Alexandrov, L.~B., Kim, J., Haradhvala, N.~J., many others, and Stratton, M.~R.
  (2020).
\newblock The repertoire of mutational signatures in human cancer.
\newblock {\em Nature}, 578(7793):94--101.

\bibitem[Alexandrov et~al., 2013]{alexandrov2013}
Alexandrov, L.~B., Nik-Zainal, S., Wedge, D.~C., Campbell, P.~J., and Stratton,
  M.~R. (2013).
\newblock Deciphering signatures of mutational processes operative in human
  cancer.
\newblock {\em Cell reports}, 3(1):246--259.

\bibitem[Arndt et~al., 2003]{ArndtBurgeHwa2003}
Arndt, P.~F., Burge, C.~B., and Hwa, T. (2003).
\newblock {DNA} sequence evolution with neighbor-dependent mutation.
\newblock {\em Journal of Computational Biology}.

\bibitem[Bertl et~al., 2018]{Bertl2018}
Bertl, J., Guo, Q., Juul, M., Besenbacher, S., Nielsen, M.~M., Hornshøj, H.,
  Pedersen, J.~S., and Hobolth, A. (2018).
\newblock A site specific model and analysis of the neutral somatic mutation
  rate in whole-genome cancer data.
\newblock {\em BMC Bioinformatics}, 19(147).

\bibitem[Biernacki et~al., 2003]{biernacki2003choosing}
Biernacki, C., Celeux, G., and Govaert, G. (2003).
\newblock Choosing starting values for the em algorithm for getting the highest
  likelihood in multivariate gaussian mixture models.
\newblock {\em Computational Statistics \& Data Analysis}, 41(3-4):561--575.

\bibitem[Cemgil, 2009]{cemgil2008bayesian}
Cemgil, A.~T. (2009).
\newblock Bayesian inference for non--negative matrix factorisation models.
\newblock {\em Computational {I}ntelligence and {N}euroscience}, Article ID
  785152.

\bibitem[Dempster et~al., 1977]{DempsterLairdRubin1977}
Dempster, A.~P., Laird, N.~M., and Rubin, D.~B. (1977).
\newblock Maximum likelihood from incomplete data via the {EM} algorithm.
\newblock {\em Journal of the Royal Statistical Society. Series B
  (Methodological)}, 39(1):1--38.

\bibitem[Du and Varadhan, 2020]{Du2020}
Du, Y. and Varadhan, R. (2020).
\newblock {SQUAREM}: An {R} package for off-the-shelf acceleration of {EM},
  {MM} and other {EM}-like monotone algorithms.
\newblock {\em Journal of Statistical Software}, 92(7):1--41.

\bibitem[Hoang et~al., 2013]{Hoang2013}
Hoang, M.~L., Chen, C.-H., Sidorenko, V.~S., He, J., Dickman, K.~G., Yun,
  B.~H., Moriya, M., Niknafs, N., Douville, C., Karchin, R., Turesky, R.~J.,
  Pu, Y.-S., Vogelstein, B., Papadopoulos, N., Grollman, A.~P., Kinzler, K.~W.,
  and Rosenquist, T.~A. (2013).
\newblock Mutational signature of aristolochic acid exposure as revealed by
  whole-exome sequencing.
\newblock {\em Science Translational Medicine}, 5(197):197--197.

\bibitem[Hobolth, 2008]{Hobolth2008}
Hobolth, A. (2008).
\newblock A {M}arkov chain {M}onte {C}arlo expectation maximization algorithm
  for statistical analysis of {DNA} sequence evolution with neighbor-dependent
  substitution rates.
\newblock {\em Journal of Computational and Graphical Statistics}, pages
  138--162.

\bibitem[Hwang and Green, 2004]{Hwang13994}
Hwang, D.~G. and Green, P. (2004).
\newblock Bayesian {M}arkov chain {M}onte {C}arlo sequence analysis reveals
  varying neutral substitution patterns in mammalian evolution.
\newblock {\em Proceedings of the National Academy of Sciences},
  101(39):13994--14001.

\bibitem[Lal et~al., 2021]{Lal2020}
Lal, A., Liu, K., Tibshirani, R., Sidow, A., and Ramazzotti, D. (2021).
\newblock De novo mutational signature discovery in tumor genomes using
  {S}parse{S}ignatures.
\newblock {\em PLoS computational biology}, 17(1):e1009119.

\bibitem[Laursen and Hobolth, 2022]{LaursenHobolth2021}
Laursen, R. and Hobolth, A. (2022).
\newblock A sampling algorithm to compute the set of feasible solutions for
  nonnegative matrix factorization with an arbitrary rank.
\newblock {\em SIAM Journal on Matrix Analysis and Applications},
  43(1):257--273.

\bibitem[Lawrence et~al., 2013]{lawrence2013}
Lawrence, M.~S., Stojanov, P., Polak, P., Kryukov, G.~V., Cibulskis, K.,
  Sivachenko, A., Carter, S.~L., Stewart, C., Mermel, C.~H., Roberts, S.~A.,
  et~al. (2013).
\newblock Mutational heterogeneity in cancer and the search for new
  cancer-associated genes.
\newblock {\em Nature}, 499(7457):214--218.

\bibitem[Lee et~al., 2017]{Lee2017}
Lee, J. Y.~L., Green, P.~J., and Ryan, L.~M. (2017).
\newblock On the '{P}oisson {T}rick' and its extensions for fitting multinomial
  regression models.
\newblock {\em arXiv: 1707.08538}.

\bibitem[Lindberg et~al., 2019]{lindberg2019intragenomic}
Lindberg, M., Bostr{\"o}m, M., Elliott, K., and Larsson, E. (2019).
\newblock Intragenomic variability and extended sequence patterns in the
  mutational signature of ultraviolet light.
\newblock {\em Proceedings of the National Academy of Sciences},
  116(41):20411--20417.

\bibitem[McCullagh and Nelder, 1989]{McCullaghNelder1989}
McCullagh, P. and Nelder, J.~A. (1989).
\newblock {\em Generalized Linear Models, 2nd edition}.
\newblock Chapman and Hall.

\bibitem[Nik-Zainal and Morganella, 2017]{Nik-Zainal2017}
Nik-Zainal, S. and Morganella, S. (2017).
\newblock Mutational signatures in breast cancer: The problem at the {DNA}
  level.
\newblock {\em Clinical Cancer Research}, 23(11):2617--2629.

\bibitem[Shen et~al., 2020]{Shen2020}
Shen, Y., Ha, W., Zeng, W., Queen, D., and Liu, L. (2020).
\newblock Exome sequencing identifies novel mutation signatures of {UV}
  radiation and trichostatin {A} in primary human keratinocytes.
\newblock {\em Scientific Reports}, 10(4943).

\bibitem[Shiraishi et~al., 2015]{shiraishi2015}
Shiraishi, Y., Tremmel, G., Miyano, S., and Stephens, M. (2015).
\newblock A simple model-based approach to inferring and visualizing cancer
  mutation signatures.
\newblock {\em PLoS {G}enetics}, 11(12):e1005657.

\bibitem[Shmueli, 2010]{Shmueli2010}
Shmueli, G. (2010).
\newblock To explain or to predict?
\newblock {\em Statistical Science}, 25(3):289--310.

\bibitem[Woo and Li, 2012]{WooLi2012}
Woo, Y.~H. and Li, W.-H. (2012).
\newblock {DNA} replication timing and selection shape the landscape of
  nucleotide variation in cancer genomes.
\newblock {\em Nature Communications}, 3(1004).

\end{thebibliography}
\end{document}